\documentclass[prl,twocolumn,superscriptaddress,nofootinbib]{revtex4-1}
\usepackage{graphicx}
\usepackage{bm}
\usepackage{amssymb}
\usepackage{color}
\usepackage{amsmath}
\usepackage{mathrsfs} 
\usepackage{amstext}
\usepackage[english]{babel}
\usepackage{latexsym}
\usepackage{array}             
\usepackage{multirow}         
\usepackage[usenames,dvipsnames]{xcolor}
\usepackage[colorlinks=true,citecolor=Blue,linkcolor=RubineRed,urlcolor=Blue]{hyperref}
\usepackage{tocloft}
\usepackage{multibib}

\newcommand{\pac}[1]{ \left\{ #1 \right\} }
\newcommand{\pap}[1]{\left( #1 \right)}
\newcommand{\pas}[1]{\left[#1 \right]}

\def\la{\langle}
\def\ra{\rangle}

\newcommand{\beq}{\begin{equation}}
\newcommand{\eeq}{\end{equation}}
\newcommand{\beqa}{\begin{eqnarray}}
\newcommand{\eeqa}{\end{eqnarray}}

\begin{document}
\title{Locality of spontaneous  symmetry breaking
and  universal  spacing distribution of topological defects formed across a phase transition}
\author{Adolfo del Campo\;\href{https://orcid.org/0000-0003-2219-2851}{\includegraphics[scale=0.45]{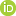}}}
\affiliation{Department  of  Physics  and  Materials  Science,  University  of  Luxembourg,  L-1511  Luxembourg, Luxembourg}
\affiliation{Donostia International Physics Center,  E-20018 San Sebasti\'an, Spain}
\author{Fernando Javier G\'omez-Ruiz\;\href{https://orcid.org/0000-0002-1855-0671}{\includegraphics[scale=0.45]{orcid}}}
\affiliation{Instituto de F\'isica Fundamental IFF-CSIC, Calle Serrano 113b, Madrid 28006, Spain}
\affiliation{Donostia International Physics Center,  E-20018 San Sebasti\'an, Spain}
\author{Hai-Qing Zhang\;\href{https://orcid.org/0000-0003-4941-7432}{\includegraphics[scale=0.45]{orcid}}}
\affiliation{Center for Gravitational Physics, Department of Space Science \& International Research Institute
for Multidisciplinary Science, Beihang University,
Beijing 100191, China}
\begin{abstract}
The crossing of a continuous phase transition results in the formation of topological defects with a density predicted by the Kibble-Zurek mechanism (KZM). We characterize the spatial distribution of point-like topological defects in the resulting nonequilibrium state and model it using a  Poisson point process in arbitrary spatial dimensions with KZM density. Numerical simulations in a one-dimensional $\phi^4$ theory unveil short-distance defect-defect corrections stemming from the kink excluded volume, while in two spatial dimensions, our model accurately describes the vortex spacing distribution in a strongly coupled superconductor indicating the suppression of defect-defect spatial correlations.\\
\\
DOI: \href{https://link.aps.org/doi/10.1103/PhysRevB.106.L140101}{10.1103/PhysRevB.106.L140101}
\end{abstract}

\maketitle
{\it Introduction.}  Spontaneous symmetry breaking in finite time leads to the formation of topological defects. The dynamics of a continuous phase transition, classical or quantum, is characterized by the breakdown of adiabaticity resulting from critical slowing down. Facing a degenerate vacuum manifold, spatially separated regions of the driven system make independent choices of the broken symmetry. Topological defects, such as domain walls in a ferromagnet or vortices in a superconductor, form at the interface between adjacent domains characterized by a homogeneous order parameter. Exploiting equilibrium scaling theory, the Kibble-Zurek mechanism (KZM) predicts the average number of spontaneously formed topological defects as a function of the quench time $\tau_Q$ in which the phase transition is crossed, e.g.,  by modulating a control parameter $\lambda$, such as the temperature or an external magnetic field, across its critical value $\lambda_c$~\cite{Kibble76a,Kibble76b,Zurek96a,Zurek96c,DZ14}.

The KZM scaling has been verified in a variety of platforms, studying soliton formation in cigar-shaped Bose-Einstein condensates~\cite{Lamporesi13} and kinks in trapped ions~\cite{EH13,Ulm13,Pyka13}, with moderate system sizes and inhomogeneous samples~\cite{DKZ13}. The scaling of the vortex density has been established in hexagonal manganites~\cite{Griffin12,Lin14}, Bose gases~\cite{Navon15,Chomaz15}, and unitary Fermi gases~\cite{Shin19}. Across quantum phase transitions, which can also be described by the KZM~\cite{Damski05,Polkovnikov05,Dziarmaga05,ZDZ05,Polkovnikov11}, the universal scaling has been probed using two-level systems~\cite{Guo14,Wang14,Wu16,Cui16,Cui19}, D-Wave machines~\cite{Weinberg20,Bando20,King22}, and Rydberg gases~\cite{Lukin17}.

A crucial ingredient of the KZM is the local character of spontaneous symmetry breaking. Kibble proposed that defects formed according to the geodesic rule, in a probabilistic fashion~\cite{Kibble76a}, a feature that has been explored by merging experimentally independent Bose-Einstein condensates~\cite{Scherer07}. This observation has motivated the unraveling of universal signatures in critical dynamics that lie beyond the scope of the KZM, e.g.,  by focusing on the full counting statistics of topological defects~\cite{Cincio07,delcampo18,Cui19,GomezRuiz20,Bando20,delCampo:2021rak,Mayo21,Goo21,King22}. 

What kind of spatial correlations between topological defects are consistent with the locality of symmetry breaking?
Pioneering works by Halperin~\cite{Halperin81} and Liu and Mazenko~\cite{LiuMazenko92} focused on growth dynamics and phase-ordering kinetics. In these studies, the characterization of correlations between topological defects was pursued using two-point correlation functions, leaving room for more stringent tests involving higher-order correlations. An important precedent of our work is the observation by Zurek that spatial defect statistics can be used to characterize spontaneously formed supercurrents~\cite{Zurek13}.

In this Research Letter, we explore the extent to which spontaneous formation of topological defects is correlated at different locations. Specifically, we focus on the spacing distribution of topological defects generated during the crossing of a continuous phase transition in finite time. We combine elements of stochastic geometry and the KZM to describe the formation of topological defects by a Poisson point process and put forward a universal prediction for the defect spacing distribution which varies with the spatial dimension. This theory is corroborated by numerical simulations of kink statistics in a one-dimensional $\phi^4$ theory and vortex formation in a two-dimensional holographic superconductor. 

{\it Universal defect spacing distribution.}  At equilibrium, the correlation length $\xi$ and the relaxation time $\tau$ diverge according to the power laws $\xi=\xi_0/|\epsilon|^{-\nu}$ and $\tau=\tau_0/|\epsilon|^{-z\nu}$ as a function of the distance $\epsilon=(\lambda-\lambda_c)/\lambda_c$ to the critical point  $\lambda_c$. Here, $\nu$  denotes the the correlation-length critical exponent, and $z$ is the dynamic critical exponent.
For a  linear quench of the form $\epsilon=t/\tau_Q$, the KZM~\cite{Kibble76a,Kibble76b,Zurek96a,Zurek96c}  sets the average distance between topological defects  equal to the KZM correlation length
\beqa
\label{hatxi}
\hat{\xi}=\xi_0\left(\frac{\tau_Q}{\tau_0}\right)^{\frac{\nu}{1+z\nu}}.
\eeqa
Thus the typical density of topological defects is
$
\rho\propto\hat{\xi}^{-d}=\xi_0^{-d}\left(\tau_0/\tau_Q\right)^{\frac{d\nu}{1+z\nu}}
$
in $d$ spatial dimensions.

We consider a Poisson point process associated with the random distribution of $N$ point like topological defects in a volume $V$ with homogeneous density $\rho$ fixed by the KZM. 
We use the well-known fact that a $(d-1)$-dimensional sphere of radius $R$ enclosing a $d$-dimensional ball $B_d(R)=\{x\in\mathbb{R}^d:\| x\|\leq R\}$ has surface $S_{d-1}(R)$ and volume $V_d(R)$ given by \cite{Huang87book}
\beqa
S_{d-1}(R)=\frac{2\pi^{d/2}}{\Gamma\left(\frac{d}{2}\right)}R^{d-1},\quad 
V_{d}(R)=\frac{\pi^{d/2}}{\Gamma\left(\frac{d}{2}+1\right)}R^{d},
\eeqa
in terms of the gamma function $\Gamma(x)$.
The distribution of $N$ topological defects in volume $V=V_{d}(R)$ leads to an effective volume occupied by each topological defect
\beqa
v_{d}=\frac{V_{d}(R)}{N}=\frac{1}{\rho}\propto\hat{\xi}^d.
\eeqa
We choose the proportionality constant such that
\beqa
V_{d}(R)=Nv_{d}=NV_{d}(\hat{\xi}),
\label{VolTot}
\eeqa
e.g., by considering each defect at the center of a $d$-dimensional ball of radius $\hat{\xi}$, which implies $R=N^{\frac{1}{d}}\hat{\xi}$.

The defect spacing distribution can be estimated  taking as a reference a given defect, and determining the probability of finding any of the other $(N-1)$ defects  at a distance between  $s$ and $s+ds$, with the remaining $(N-2)$ defects being located farther away, 
\beqa
P(s)ds&=&(N-1)\frac{S_{d-1}(s)ds}{V_{d}(R)}\left[1-\frac{V_{d}(s)}{V_{d}(R)}\right]^{N-2}.
\eeqa
We note that the above expression for $P(s)$  fulfills the normalization condition
$\int_0^RP(s)ds=1$.
This distribution generalizes the familiar distribution for a Poissonian random process on a plane, studied, e.g., in the context of random matrix theory.
Making use of (\ref{VolTot}) and introducing the dimensionless scaled spacing
$
X=s/R=s/(N^{\frac{1}{d}}\hat{\xi}),
$
one finds the spacing distribution of topological defects
$
P(s)ds=d(N-1)X^{d-1}\left[1-X^d\right]^{N-2}dX.
$
\begin{figure}[t]
\centering
\includegraphics[width=1\linewidth]{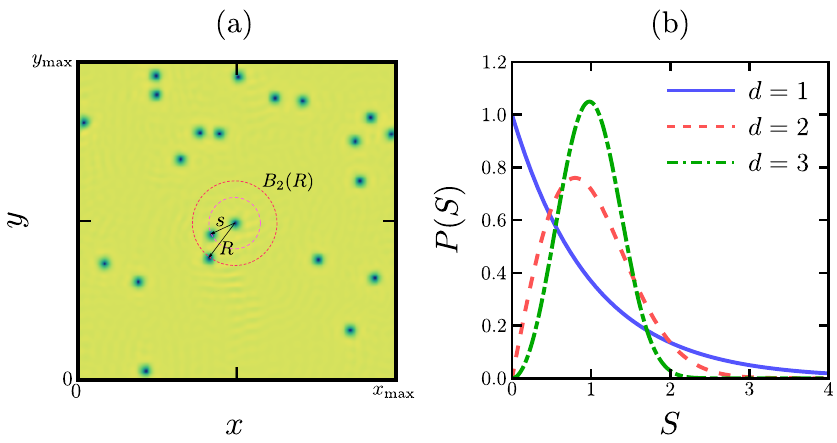}
\caption{\label{PsPDF}  Universal defect spacing distribution. (a) Schematic representation of $N$ point like topological defects enclosed in a  two-dimensional ball $B_2\pap{R}$ with radius $R$. (b)  Distribution $P(S)$ of the spacing between topological defects as a function of the dimensionless spacing normalized to the mean, $S=sd/[\hat{\xi}\Gamma\left(\frac{1}{d}\right)]$, where $s$ is the shortest distance between defects, $d$ denotes the spatial dimension, and $\hat{\xi}$ is the non-equilibrium correlation length predicted by the KZM, which exhibits the universal power-law scaling with the quench time in Eq. (\ref{hatxi}).
In one dimension the distribution is exponential, while $P(S)$ takes the form of a Wigner-Dyson distribution in $d=2$. }
\end{figure}
We shall focus on the large $N$ limit, in which the normalized distribution reads
\beqa
P(s)ds=dNX^{d-1}e^{-NX^d}dX.
\eeqa
The mean spacing, $\la s\ra=\int_0^RP(s)s ds$, is given by
\beqa
\la s\ra=d N^{1+\frac{1}{d}}\hat{\xi}\int_0^1X^{d}e^{-NX^d}dX\nonumber\approx\frac{\hat{\xi}}{d}\Gamma\left(\frac{1}{d}\right).
\eeqa
with the  correlation length  $\hat{\xi}$ in Eq.\eqref{hatxi} set by the KZM.
To bring out the universal character of $P(s)$ with the quench time, it proves convenient to
introduce the dimensionless defect spacing in units of the mean
$S=s/\la s\ra$. Using it, the normalized defect spacing distribution in the large-$N$ limit takes the form
\beqa
P(S)=dr^dS^{d-1}e^{-r^dS^d},
\quad 
r=\frac{\la s\ra}{\hat{\xi}}=\frac{1}{d}\Gamma\left(\frac{1}{d}\right), 
\eeqa
which is thus independent of the quench time $\tau_Q$. This distribution is normalized in the domain $S\in[0,\infty)$, has unit mean $\la S\ra=\int_0^\infty SP(S)dS=1$, and has a maximum at
$S_{\rm max}=(1-\frac{1}{d})^d/\Gamma\left(1+\frac{1}{d}\right)$ with value $ P(S_{\rm max})=(d-1)^{\frac{d-1}{d}} d^{\frac{1}{d}} e^{\frac{1}{d}-1}\Gamma\left(1+\frac{1}{d}\right).
$ In addition, the fluctuations in the defect spacing as quantified by the variance read
$ \Delta S^2=\Gamma\left(\frac{2+d}{d}\right)/\Gamma\left(1+\frac{1}{d}\right)^2-1$,
diminishing with increasing spatial dimension $d$, and vanishing as $d\rightarrow\infty$. Specifically, we note that for spatial dimensions $d=1,2,3$, $r=1,\sqrt{\pi}/2,\frac{1}{3}\Gamma(\frac{1}{3})$, and
\beqa
d=1,\quad P(S)&=&e^{-S},\\
d=2,\quad P(S)&=&\frac{\pi}{2}Se^{-\frac{\pi}{4}S^2},\label{PS2}\\
d=3,\quad P(S)&=&\frac{1}{9}\Gamma\left(\frac{1}{3}\right)^3S^2e^{-\frac{1}{27}\Gamma\left(\frac{1}{3}\right)^3S^3},
\eeqa
with fluctuations $\Delta S^2\approx 1, 0.27,$ and $0.13$, respectively. 
These distributions are plotted in Fig.~\ref{PsPDF}. For $d=1$ the distribution is exponential, familiar in the context of the energy-level spacing of integrable systems. For $d=2$, the defect spacing distribution is given by the well-known Wigner-Dyson distribution, familiar from random matrix theory.
For arbitrary $d$, it is known as the Brody distribution in the context of quantum chaos \cite{Brody73,Brody81}.   
\begin{figure}[t]
\centering
\includegraphics[width=1\linewidth]{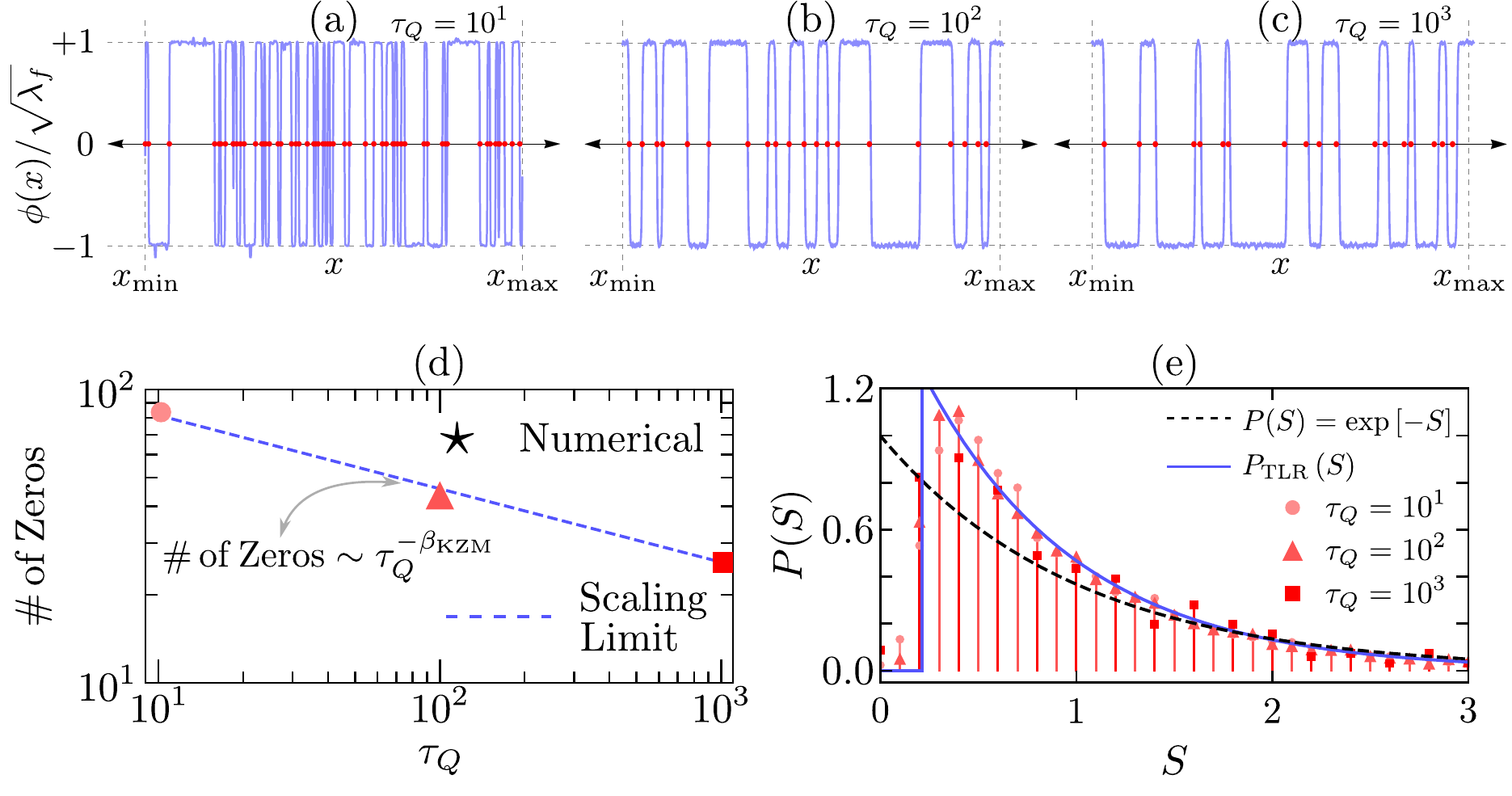}
\caption{\label{PsPDF1D} {\bf Statistics of  kinks spontaneously formed in a one-dimensional $\phi^4$ theory.} The top panels (a) (c) A typical nonequilibrium field configuration $\phi\pap{x}$ supporting kinks is shown, when the transition is crossed at different quench times $\tau_Q = 10^1$, $10^2$, and $10^3$. The position of every kink is represented by a red dot over $\phi\pap{x}=0$. (d) Scaling of the number of zeros as a function of the quench time. A fitting to the  scaling of the average number of zeros with the quench time of the form (number of zeros $\propto \tau_Q^{-\beta_{\rm KZM}}$) yields a value of the  power-law exponent  $\beta_{{\rm KZM}}=0.248\pm  0.004$ in agreement with the KZM prediction $d\nu/(1+z\nu)=1/4$ for $d=1$, $\nu=1/2$, and $z=2$. (e) Distribution $P(S)$ of the spacing between nearest-neighbor topological defects. While the details of the distribution are well reproduced by an exponential decay, there are strong short-distance suppression of $P(S)$ and bunching at a spacing comparable to the mean, $S\sim 1$.   Deviations are captured by the TLR distribution with $\phi=0.11$. For each value of $\tau_Q$, data are collected from 10 000 independent trajectories. } 
\end{figure}

Point random processes assume no correlations between defects beyond the fact that the density is homogeneous throughout the system. As a result, they provide a  structureless model,  a natural reference one in the context of the KZM. We next explore the extent to which the spontaneous formation of topological defects can be described
by point random processes with KZM density by considering some paradigmatic models. 

{\it Example 1: Spontaneous $\mathbb{Z}_2$ breaking in a one-dimensional system.---}
Let us consider a paradigmatic scenario of spontaneous symmetry breaking to explore the defect-defect spacing distribution following the crossing of a phase transition in finite time: a Ginzburg-Landau theory with real field $\phi$ undergoing Langevin dynamics. The validity of the KZM in this setting has been reported in Ref.~\cite{Laguna98,delcampo10,Nigmatullin16}. In addition, the universality of the number distribution of topological defects, beyond the KZM, has also been established  in Ref.~\cite{GomezRuiz20}. We consider a generic  Ginzburg-Landau potential given by $V\pap{\phi}=\frac{1}{8}\pap{\phi^4 - 2\epsilon\phi^2+1}$. The parameter $\epsilon$ measures the distance from the phase transition.
Minimizing  $V\pap{\phi}$ with respect to $\phi$ gives two possible minima given by $\phi_{\min} \pap{t}= \pac{0, \pm\sqrt{\epsilon\pap{t}}}$. A second order phase transition occurs if $\epsilon$ changes its sign. When $\epsilon<0$, the field $\phi$ fluctuates around its expectation value $\langle \phi \rangle =0$. However, when $\epsilon>0$, the field $\phi$ settles locally around one of two minima $\langle \phi \rangle=\pm\sqrt{\epsilon}$.
We consider a quench described by  
\begin{equation}\label{linearamp}
\epsilon \pap{t}=\lambda_{i} + \frac{t}{\tau_Q}\left|\lambda_{f}-\lambda_{i}\right|. 
\end{equation}
 The system is in contact with a thermal reservoir and obeys the Langevin equation
\begin{equation}\label{LangEq}
\ddot{\phi}+\eta\dot{\phi}-\partial_{xx}\phi+\partial_{\phi} V\pap{\phi}=\xi\pap{x,t},
\end{equation}
where $\eta >0$ is a global  damping constant. The noise $\xi\pap{x,t}$ is a real Gaussian process with zero mean and autocorrelation function  $\langle \xi\pap{x,t}\xi\pap{x',t'}\rangle = 2\eta T\delta\pap{x-x'}\delta\pap{t-t'}$. Details of the numerical integration to determine trajectories of the Ginzburg-Landau real field $\phi$ can be found in the Supplemental Material~\cite{*[{See Supplemental Material at }] [{for details of the calculations and derivations.}] SM}.
We  simulate the dynamics in the overdamped regime and fix the temperature $T=0.01$ and $\eta=1$ (the same regime of parameters used in Ref.~\cite{Laguna98}). The stochastic evolution is described  using a computational numerical grid from $x_{\min}=0$ to $x_{\max}=500$ given by $5000$ partitions. The correlation length critical exponent takes the mean-field value $\nu=1/2$. In  the overdamped regime, the dynamic critical exponent is $z=2$. The KZM prediction for the typical size of the domains is thus $\hat{\xi}=\xi_0(\tau_Q/\tau_0)^{\frac{1}{4}}$, numerically confirmed in  Ref.\cite{Laguna98,GomezRuiz20}. 

The distribution $P(S)$ of the spacing between adjacent topological defects, normalized to the mean, is shown in Fig.~\ref{PsPDF1D}.  The tail of distribution for $S>1.5$ is well described by the exponential function $\exp(-S)$. However, important deviations are manifested for smaller values of $S$. In particular, $P(S)$ is highly suppressed for small values of $S$ while an enhancement of the probability for $S\simeq 1$ is observed.  This is to be expected as one dimension enhances correlations between topological defects. Furthermore, in one spatial dimension, a kink can only be surrounded by antikinks. Coarsening by the annihilation of kink and anti-kink pairs is thus unavoidable even in the KZM scaling regime of the dynamics for small lattice spacing. 
In addition, topological defects such as kinks are actually not point-like and have a finite healing length.

One may wonder whether the distribution can be reproduced by considering the excluded volume of each defect. To this end, we consider the spacing between randomly distributed disks in one spatial dimension, characterized by Torquato, Lu, and Rubinstein (TLR). In terms of 
 the packing fraction $\phi=N\sigma/L\in [0,1]$ for $N$ topological defects of  radius $\sigma$, the TLR spacing distribution  reads~\cite{TLR90,Torquato95}
\beqa
P_{\rm TLR}(S)=\frac{1+\phi}{1-\phi}\exp\left[-\frac{1+\phi}{1-\phi}S+\frac{2\phi}{1-\phi}\right],
\eeqa
 and vanishes identically for small spacing values when  $S<2\phi/(1+\phi)$. The comparison with this distribution is shown as well in Fig.~\ref{PsPDF1D}. The TLR distribution captures accurately the corrections to the ideal distribution that arise at all distances from the excluded volume, with a weak dependence of the optimal packing fraction $\phi$ on the quench time; see Supplemental Material~\cite{SM} for additional details. 

{\it Example 2: Spontaneous ${\rm U}(1)$ breaking in a two-dimensional strongly-coupled superconductor.} We next turn our attention to an example in one higher spatial dimension and characterize the defect spacing distribution of spontaneously formed vortices in a newborn strongly-coupled superconductor. The latter can be described using holographic duality which makes use of a gravitational theory with one additional space-like dimension \cite{Hartnoll18book}. 
Holographic quantum matter provides a natural setting to explore the role of strong coupling in topological defect formation and has been used to establish the validity of the KZM in this regime, up to subleading corrections in the quench time \cite{Sonner:2014tca,Chesler:2014gya,Li:2019oyz,Zeng:2019yhi,delCampo:2021rak,Li:2021iph,Li:2021dwp}.

The setup and numerical scheme we adopt are described in detail in the Supplemental Material~\cite{SM}, which includes  Refs.~\cite{Trefethen00book,Skenderis02,Domenech10,CheslerYaffe14}. 
We consider the paradigmatic Lagrangian of the Maxwell-complex scalar model for a holographic superconductor \cite{Hartnoll:2008vx},
\begin{equation}\label{lag}
\mathcal{L} = -\frac{1}{4} F_{\mu \nu} F^{\mu \nu} - |D \Psi|^2 - m^2 |\Psi|^2.
\end{equation}
where $D=\nabla -iA$ with $A$ being the U(1) gauge field and $\Psi$ being the complex scalar field (using  units with $e=c=\hbar=k_B=1$). The equations of motion read $D_\mu D^\mu\Psi-m^2\Psi=0, \nabla_\mu F^{\mu\nu}=i\left(\Psi^* D^\nu\Psi-\Psi{(D^\nu\Psi)^*}\right)$. The ansatz we  take is $\Psi = \Psi(t,z,x,y)$, $A_t = A_t(t,z,x,y)$, $A_x = A_x(t,z,x,y)$, $A_y = A_y(t,z,x,y)$, and $A_z = 0$. In the simulation, we set the size of the system as $(x,y)=(50,50)$ and the mass square of the scalar field as $m^2=-2$.
Numerical simulations and theoretical analysis indicate that in this system the correlation-length critical exponent $\nu=1/2$ while the dynamic critical exponent $z=2$ \cite{Zeng:2019yhi,delCampo:2021rak}, so that the KZM correlation length is consistent with $\hat{\xi}=\xi_0(\tau_Q/\tau_0)^{\frac{1}{4}}$.

 Before quenching,  the system is thermalized by adding the noise $\xi(x_i,t)$ into the bulk and satisfying the conditions $\langle \xi(x_i,t)\rangle=0$ and $\langle \xi(x_i,t)\xi(x_j,t')\rangle=h\delta(t-t')\delta(x_i-x_j)$, with a small amplitude $h=10^{-3}$. Subsequently, a linear quench of the temperature, from $T_i = 1.4T_c$ to $T_f = 0.8T_c$, prompts the system to evolve from a normal metal state to a superconductor state.  As a result of the U(1) symmetry breaking, vortices form spontaneously as predicted by the KZM. 

\begin{figure}[t]
\centering
\includegraphics[width=1\linewidth]{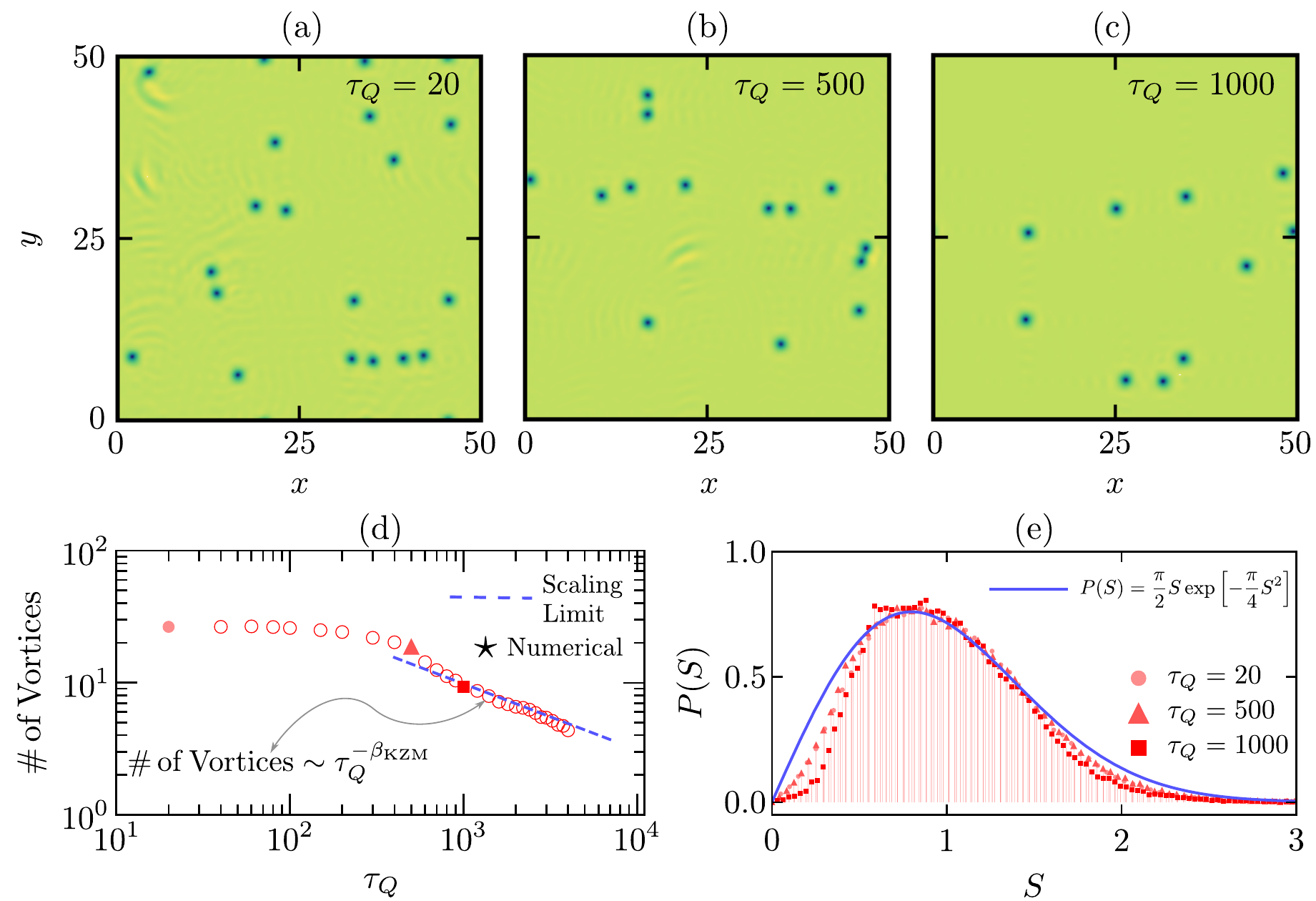}
\caption{{\bf  Statistics of vortices in a newborn strongly coupled superconductor.}  (a)-(c) Spontaneously formed vortices in a newborn holographic superconductor are shown in density plots  of the order parameter for three different quench rates. The average vortex number decreases with the quench rates. (d) The dependence of the average vortex number on the quench time is consistent with a power-law scaling, with fitted value $\beta_{\rm KZM}=0.512\pm 0.005$ in agreement with the prediction by the KZM with $d\nu/(1+z\nu)=1/2$ for $d=2$,  $\nu=1/2$, and $z=2$. (e)  Spacing distribution $P(S)$ for three different quenches. The numerical data are consistent with the theoretical fitting line $P(S)= \frac{\pi}{2}S e^{-\frac{\pi}{4}S^2}$. For each value of $\tau_Q$, data are collected from 10 000 independent trajectories.}
\label{pic3}
\end{figure}

We evaluate the spatial distribution of vortices until the system enters the equilibrium state. Typical spatial distributions of the vortices are shown in Fig.~\ref{pic3}(a)-\ref{pic3}(a) for three different kinds of quench rates, $\tau_Q=20, 500$, and $1000$. The number of vortices decreases with the quench time. The scaling of the average number of vortices as a function of the quench time is shown in the down left panel of Fig. \ref{pic3}. For slow quenches, the scaling law satisfies the KZM with the power $\beta_{\rm KZM}\approx1/2$ in 2D.  The corresponding vortex spacing distributions are shown in Fig.~\ref{pic3}(e).  In obtaining Fig.~\ref{pic3}, we first choose a vortex as a reference and then determine the closest vortex to it, regardless of its topological charge. Because we used the lattice square to simulate the system, we regard the vortices sitting at the same circumference of a square to have the same distance to the centering vortex. The distributions of $P(S)$ estimated from $10^4$ realizations are in good agreement with the theoretical prediction Eq. \eqref{PS2} in two spatial dimensions. Numerically estimated distributions are slightly narrower than the corresponding Wigner-Dyson distribution. In two dimensions the interactions between topological defects are logarithmic with the distance between topological defects~\cite{Nelson02book}, excluding the validity of a short-range hard-disk model as shown in Supplemental Material~\cite{SM}. 
As the deviations from the ideal distribution are more pronounced for slower quenches, we attribute them to coarsening.   

{\it Conclusions.} Combining the scaling theory of continuous phase transitions, the Kibble-Zurek mechanism, and elements of stochastic geometry, we have described the spontaneous formation of topological defects across a continuous phase transition as a Poisson point process with KZM density. Using this framework, we have predicted the universal form of the defect spacing distribution as a function of the quench time and spatial dimension. 
In one-dimensional systems, defect-defect correlations are enhanced and can be taken into account by considering the finite size of defects. The theory is expected to accurately reproduce the spacing distribution in higher dimensions, as we have shown in a two-dimensional setting, where remaining deviations are attributed to coarsening. 
Our results are amenable to experimental test with established technology,  exploiting any of the platforms previously used to probe KZM scaling, provided it is endowed with spatial resolution, as is the case with trapped-ion systems \cite{Ulm13,Pyka13}, colloidal monolayers \cite{Keim15}, multiferroics \cite{Griffin12,Lin14}, ultracold gases in various geometries \cite{Lamporesi13,Navon15,Chomaz15,Shin19}, and quantum simulators, to name some prominent examples.

{\it Acknowledgements.} The authors are indebted to Bogdan Damski for useful comments and suggestions. FJG-R acknowledges financial support from European Commission FET-Open project AVaQus GA 899561. HQZ was partially supported by the National Natural Science Foundation of China (Grants No. 11875095 and 12175008).
\bibliography{defects_spacing_Bib}
\newpage
\pagebreak
\clearpage
\widetext
\setcounter{equation}{0}
\setcounter{figure}{0}
\setcounter{table}{0}
\setcounter{section}{0}
\setcounter{page}{1}
\makeatletter
\renewcommand{\theequation}{S\arabic{equation}}
\renewcommand{\thefigure}{S\arabic{figure}}
\renewcommand{\bibnumfmt}[1]{[S#1]}
\renewcommand{\citenumfont}[1]{#1}
\begin{center}
\textbf{\large ---Supplemental Material---\\
 Locality of spontaneous  symmetry breaking and universal spacing distribution of topological defects formed across a phase transition}\\
\vspace{0.5cm}
Adolfo del Campo$^{1,2}$, Fernando Javier G\'omez-Ruiz$^{3,2}$, and Hai-Qing Zhang$^{4}$\\
\vspace{0.2cm}
$^{1}${\it Department of Physics and Materials Science, University of Luxembourg, L-1511 Luxembourg, Luxembourg}\\
$^2${\it Donostia International Physics Center,  E-20018 San Sebasti\'an, Spain}\\
$^4${\it Instituto de F\'isica Fundamental IFF-CSIC, Calle Serrano 113b, Madrid 28006, Spain}\\
$^5${\it Center for Gravitational Physics, Department of Space Science \& International Research Institute for Multidisciplinary Science, Beihang University, Beijing 100191, China}
\end{center}
\section{Excluded volume by topological defects}

Consider a one-dimensional gas composed of $N$ hard rods of diameter $\sigma$ gas confined  in a line of length $L$. 
The exact (normalized) spacing distribution between particles, as quoted by Torquato-Lu-Rubinstein \cite{TLR90} in the dimensionless variable $x=s/\sigma$, reads
\beqa
P(x)=\frac{2\phi}{1-\phi}\exp\left[-\frac{2\phi}{1-\phi}(x-1)\right]\Theta(x-1),
\eeqa
in terms of the packing fraction $\phi=N\sigma/L\in [0,1]$ and the Heaviside function $\Theta(x)$. The mean spacing is
\beqa
\la x\ra =\int
P(x)xdx=\frac{1+\phi}{2\phi}.
\eeqa
In two spatial dimensions, the exact distribution is unknown.  An approximate normalized spacing distribution for hard disks has been parameterized by Torquato \cite{Torquato95}. It is given by
\beqa
\label{Torquato2d}
P(x)=8\phi(a_0x+a_1)\exp[-\phi\left(4a_0(x^2-1)+8a_1(x-1)\right)]\Theta(x-1),
\eeqa
where
\beqa
a_0=\frac{1+0.128\phi}{(1-\phi)^2},\quad a_1=\frac{-0.564\phi}{(1-\phi)^2}.
\eeqa
Here $\phi=Nv_2(\sigma/2)/v_2(R)=N\sigma^2/(4R^2)$.
Its mean reads
\beqa
\la x\ra =\int
P(x)xdx=1+\frac{\sqrt{\pi }
   e^{\frac{4 \left(a_0+a_1\right){}^2 \phi }{a_0}} \sqrt{a_0 \phi }}{4
   a_0 \phi }-\frac{\sqrt{\pi } e^{\frac{4 \left(a_0+a_1\right){}^2 \phi }{a_0}}
   \text{erf}\left(\frac{2 \left(a_0+a_1\right) \sqrt{\phi
   }}{\sqrt{a_0}}\right)}{4 \sqrt{a_0} \sqrt{\phi }}.
\eeqa
For completeness, we note that approximations to the spacing distribution in three-spatial dimensions have also been reported \cite{TLR90}. These results are useful in estimating the spacing distribution between spontaneously formed topological defects.
Imposing $P(x)dx=P(S)dS$ we find
\beqa
P(S)&=&P\left(x=\frac{s}{\sigma}\right)\frac{\la s\ra}{\sigma}\\
&=&P\left(x=\frac{S\la s\ra}{\sigma}\right)\frac{\la s\ra}{\sigma}
\\
&=&P\left(x=S\la x\ra\right)\la x\ra.
\eeqa

\begin{figure}[t]
\centering
\includegraphics[width=0.9\linewidth]{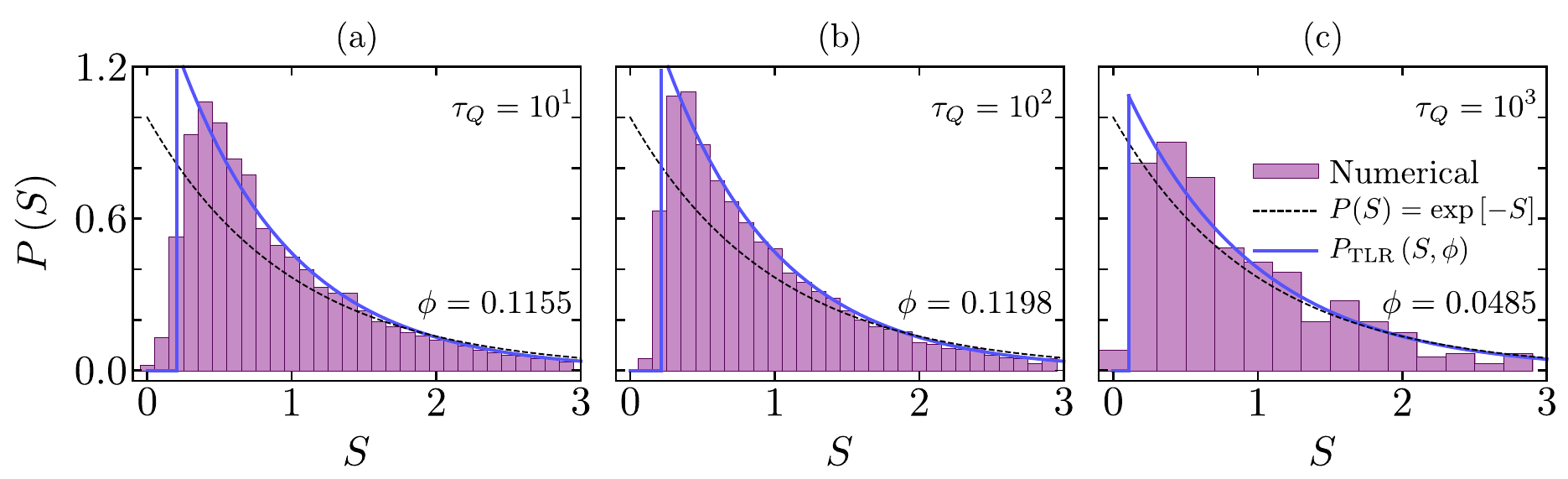}
\caption{\label{SMfigS1}{\bf Defect spacing distribution in the 1D $\phi^4$ theory and the optimal TLR distribution.} The numerical histogram for the spacing distribution is compared with the 1D ideal distribution and the optimal TRL distribution for each of the quench values considered in Fig. 1 of the main text.
}
\end{figure}

\begin{figure}[h!]
\centering
\includegraphics[width=0.9\linewidth]{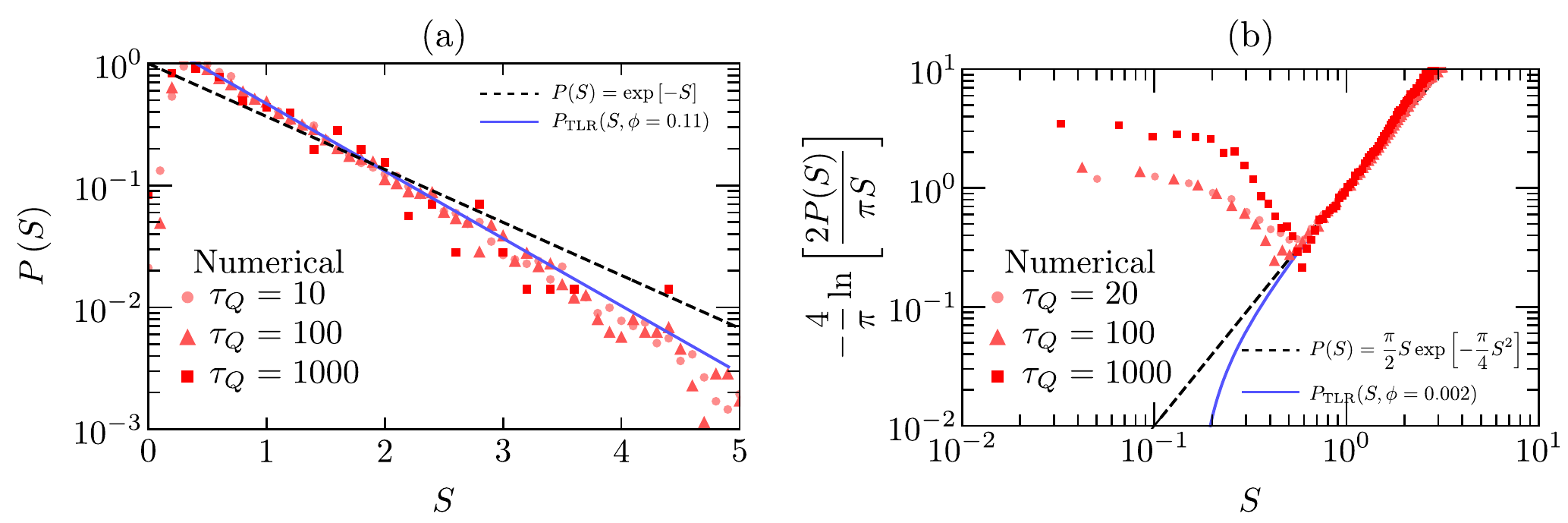}
\caption{\label{SMfigS2}{\bf Limits to homogeneous Poisson point processes in modeling the spacing distribution in the 1D $\phi^4$ theory and the holographic superconductor.} (a) For the 1D $\phi^4$ theory, the TLR distribution is shown to accurately capture the nature of the deviations from the homogeneous Poisson point process over the complete range of spacing values. The data used in Fig. 2 of the main text is shown here is a semi-log plot. (b) The numerically-generated histogram for the vortex-vortex spacing distribution in a holographic superconductor is compared with the 2D ideal distribution and the optimal TRL distribution.
The validity of both models breaks down at small values of $S$. The data used in Fig. 3 of the main text is shown here is a log-log plot.
}
\end{figure}

Thus, in 1D,
\beqa
P(S)=\frac{1+\phi}{1-\phi}\exp\left[-\frac{1+\phi}{1-\phi}S+\frac{2\phi}{1-\phi}\right]\Theta\left[S\left(1+\frac{1}{\phi}\right)-2\right],
\eeqa
used in the main text.
The packing fraction can be used as a fitting parameter. As argued, the dependence on the quench time is weak.
Figure \ref{SMfigS1} shows the histograms for each of the quench time values considered in the text and the corresponding TLR distribution with optimal $\phi$. For moderate quench times, the value of the fitted packing fraction is $\phi\approx 0.11$ while lower values are obtained in the limit of slow quench times. An alternative presentation with a semi-log scale in Fig. \ref{SMfigS2} (left panel) shows that the TLR provides an accurate description of the numerical data for all values of the normalized spacing $S$.  
%
Regarding the $d=2$ case, it has been shown in the main manuscript that small deviations appear with respect to the homogeneous Poisson point process in the numerical simulations for the holographic superconductor.
An attempt to capture these deviations with the hard-disk model in Eq. (\ref{Torquato2d}) leads to a qualitatively wrong behavior. This is consistent with the fact that interactions between vortices in two spatial dimensions are not short-range but logarithmic with the distance.   
\section{Basic setup of one-dimensional Ginzburg-Landau Potential}
In this section, we provide additional details  regarding the numerical simulation of the time evolution of a 1D real scalar field $\phi$. We consider a generic  Ginzburg-Landau potential given by~\cite{Laguna98}
\begin{equation}\label{PotV_GL}
V\pap{\phi}=\frac{1}{8}\pap{\phi^4 - 2\epsilon\phi^2+1}.
\end{equation}
The parameter $\epsilon$ measures the distance from the phase transition and varies according to:
\begin{equation}\label{linearamp2}
\epsilon \pap{t}=\lambda_{i} + \frac{t}{\tau_Q}\left|\lambda_{f}-\lambda_{i}\right|. 
\end{equation}
For the numerical simulations presented in this work, we used a fixed value of $\lambda_{i}=-2$. However, the value of $\lambda_{f}>0$ has been fixed for saturation of the the average number of zeros, as analogously done in \cite{Laguna98}. Specifically, we set $\lambda_{f}=20$ for $\tau_Q=10$ and $\lambda_{f}=4$ for $\tau_Q=100, 1000$.
Minimizing  $V\pap{\phi}$ with respect to $\phi$ gives two possible minima given by
\begin{equation}
\phi_{{\rm Min}} = \begin{cases}
0,\\
\pm\sqrt{\epsilon}.
\end{cases}
\end{equation}
A second order phase transition occurs if $\epsilon$ changes its sign. When $\epsilon<0$, the field $\phi$ fluctuates around its expectation value $\langle \phi \rangle =0$. However, when $\epsilon>0$, the field $\phi$ settles locally around one of two alternatives $\langle \phi \rangle=\pm\sqrt{\epsilon}$. The coupling of the field to a thermal reservoir is described by the Langevin equation
\begin{equation}\label{SLangEq}
\ddot{\phi}+\eta\dot{\phi}-\partial_{xx}\phi+\partial_{\phi} V\pap{\phi}=\xi\pap{x,t},
\end{equation}
where $\eta >0$ is a damping constant. The noise $\xi\pap{x,t}$ is a real Gaussian process with zero mean. The fluctuation-dissipation sets the two-point correlator $\langle \xi\pap{x,t},\xi\pap{x',t'}\rangle = 2\eta T\delta\pap{x-x'}\delta\pap{t-t'}$.\\
\\
We solve the stochastic partial differential equation \eqref{SLangEq} in a one-dimensional box, with  $x$ ranging from $x_{0}$ to a maximum value $x_{f}$ and $t$ varying from $t_{0}=0$ to $t_{f}=\tau_Q$. We divide the interval $[0, \tau_Q]$ into $N_t$ equally spaced intervals with length $\Delta t$, and the interval $[x_0,x_{f}]$ into  $N_x$ equally spaced intervals with length $\Delta x$. Specifically, we choose $x_f=500$ and $N_x=5000$.\\
\\
We seek an approximation to the true values of $\phi\pap{x,t}$ at the $ \pap{N_x \times 1}\times \pap{N_t \times 1}$ grid points. Let $\phi\pap{x_{n},t_{m}}$ denote our approximation at the grid point where $x_n=x_{0}+n \Delta x$ and $t_m=t_{0}+m \Delta t$. We use a finite difference method  to approximate the partial derivatives of $\phi$ at each grid point by difference expressions in the as yet unknown $\phi\pap{x_n,t_m}$.\\ 
We calculate $\phi\pap{x_n,t_{0}}$ for each $x_n$ directly from the initial value condition $\phi_0$ and $\dot{\phi}_0$.  
\begin{align}\label{InitialCond}
\phi_0 = \langle\phi\pap{x,t_0}\rangle=\xi_1\pap{x,t_0},&&\dot{\phi}_0 = \langle\dot{\phi}\pap{x,t_0}\rangle=\xi_2\pap{x,t_0},
\end{align}
where  $\xi_1\pap{x,t_0}$ and $\xi_2\pap{x,t_0}$ are a real Gaussian process with zero mean. Focusing on an arbitrary internal grid-point $\pap{x_n,t_m}$, and approximating derivatives by central differences both in space and time we obtain:
\begin{itemize}
\item {\bf Central Differences in Space:}
\begin{equation}
\begin{split}
\phi\pap{x+\Delta x,t}&=\phi\pap{x,t}+\Delta x \frac{\partial \phi \pap{x,t}}{\partial x}+\frac{\Delta x^2}{2!}\frac{\partial^2 \phi \pap{x,t}}{\partial x^2}+\frac{\Delta x^3}{3!}\frac{\partial^3 \phi \pap{x,t}}{\partial x^3}+\frac{\Delta x^4}{4!}\frac{\partial^4 \phi \pap{x,t}}{\partial x^4}+\ldots\\
\phi\pap{x-\Delta x,t}&=\phi\pap{x,t}-\Delta x \frac{\partial \phi \pap{x,t}}{\partial x}+\frac{\Delta x^2}{2!}\frac{\partial^2 \phi \pap{x,t}}{\partial x^2}-\frac{\Delta x^3}{3!}\frac{\partial^3 \phi \pap{x,t}}{\partial x^3}+\frac{\Delta x^4}{4!}\frac{\partial^4 \phi \pap{x,t}}{\partial x^4}+\ldots
\end{split}
\end{equation}
As a result,
\begin{equation}\label{2dx}
\frac{\phi\pap{x+\Delta x,t}-2\phi\pap{x,t}+\phi\pap{x-\Delta x,t}}{\Delta x^2}=\frac{\partial^2 \phi \pap{x,t}}{\partial x^2}+\mathcal{O}\pap{\Delta x^2}.
\end{equation}
\end{itemize}
We rewrite the Eq.~\eqref{2dx} as
\begin{equation}\label{2dx}
\frac{\partial^2}{\partial x^2} \phi_{n}^{m} = \frac{1}{\Delta x^2}
\pas{\phi_{n}^{m+1}-2\phi_{n}^{m}+\phi_{n}^{m-1}}.
\end{equation}
Therefore, the Langevin equation take the form
\begin{equation}\label{LGDiscret1}
\frac{\partial^{2}}{\partial t^2}\phi_{n}^{m}+\eta \frac{\partial}{\partial t}\phi_{n}^{m}- \frac{1}{\Delta x^2}
\pas{\phi_{n}^{m+1}-2\phi_{n}^{m}+\phi_{n}^{m-1}}+\frac{1}{2}\pas{\pap{\phi_n^m}^3-\epsilon\pap{t}\phi_{n}^{m}}=\epsilon_{n}^{m}.
\end{equation}
We used a periodical boundary condition such as: 
\begin{align}
\phi_{n}^{-1} = \phi_{n}^{N_x}, && \phi_{n}^{N_x +1} = \phi_{n}^{1}.
\end{align}
Additionally, we fixed the initial conditions 
\[
\phi_0^{m} = \langle\phi\pap{x,t_0}\rangle=\dot{\phi}_0^{m} = \langle\dot{\phi}\pap{x,t_0}\rangle=2 T \eta\xi_n^m.
  \]
We are interested in simulating the dynamics in the overdamped regime. We thus fix the temperature $T=0.01$ and $\eta=1$, choosing the same regime of parameters used in~\cite{Laguna98}. 
\section{Modeling of  a strongly-coupled superconductor}

\subsection{Equations of motion}
As we mentioned in the main text, the Lagrangian we adopt is the Maxwell-complex scalar model for a holographic superconductor~\cite{Hartnoll08},
\begin{equation}\label{lag}
\mathcal{L} = -\frac{1}{4} F_{\mu \nu} F^{\mu \nu} - |D \Psi|^2 - m^2 |\Psi|^2.
\end{equation}
where $F_{\mu\nu}=\partial_\mu A_\nu-\partial_\nu A_\mu$,  $D_\mu=\nabla_\mu -iA_\mu$ with $A_\mu$ the U(1) gauge field and $\Psi$ the complex scalar field, respectively (we work in the units with $e=c=\hbar=k_B=1$). The equations of motion read 
\begin{eqnarray}
D_\mu D^\mu\Psi-m^2\Psi=0, ~~~~\nabla_\mu F^{\mu\nu}=i\left(\Psi^* D^\nu\Psi-\Psi{(D^\nu\Psi)^*}\right).
\end{eqnarray} 
In order to describe the holographic KZM in a $(2+1)$-dimensional system, we adopt the $(3+1)$-dimensional black brane background in Eddington-Finkelstein coordinates with the line-element 
\begin{eqnarray}
{\rm d}s^2 = \frac{1}{z^2} \left(-f(z) {\rm d}t^2 - 2{\rm d}t{\rm d}z + {\rm d}x^2 + {\rm d}y^2\right),
\end{eqnarray}
where $f = 1 - (z/z_h)^3$ and the radius of AdS spacetime has been scaled  to unit one. The coordinates $(x,y)$ are the spatial directions and $z$ is the radial direction in the bulk with $z_h$ the horizon location. Therefore, the  defects we consider  live in the two-dimensional space in the boundary of the AdS spacetime, spanned by $(x,y)$ coordinates. 
We impose the ansatz by imposing $A_z=0$ and other fields are functions of the coordinates $(t,z,x,y)$, i.e.,  $\Psi = \Psi(t,z,x,y)$ and $A_m = A_m(t,z,x,y), (m=t,x,y)$. In the probe limit, the explicit form of the equations of motion for $\Psi$ (we have scaled $\Psi=\Phi z$) and $A_\mu$ read
\begin{eqnarray}
\label{eompsi}
\partial_t \partial_z \Phi - i A_t \partial_z \Phi - \frac12 [ i \partial_z A_t \Phi +\partial_z(f\partial_z\Phi)  - z \Phi 
+ \square\Phi - i ({\bf\partial\cdot A}) \Phi - (A_x^2 + A_y^2) \Phi - 2 i (A_x \partial_x \Phi + A_y \partial_y \Phi) ] = 0,&
\\
\label{eom2}
\partial_t \partial_z A_t - \square A_t - f \partial_z ({\bf\partial\cdot A}) + \partial_t ({\bf\partial\cdot A}) 
+ 2 A_t |\Phi|^2 +2f\Im(\Phi^* \partial_z \Phi)-2\Im(\Phi^* \partial_t \Phi)   = 0,&
\\
\label{eom3}
\partial_t \partial_z A_x - \frac12 \left[ \partial_z (\partial_x A_t + f \partial_z A_x) + \partial_y (\partial_y A_x - \partial_x A_y)+2\Im(\Phi^* \partial_x \Phi )  - 2 A_x |\Phi|^2 \right] = 0,&
\\
\label{eom4}
\partial_t \partial_z A_y - \frac12 \left[ \partial_z (\partial_y A_t + f \partial_z A_y) + \partial_x (\partial_x A_y - \partial_y A_x) +2\Im(\Phi^* \partial_y \Phi) - 2 A_y |\Phi|^2 \right] = 0,&
\\
\label{eom1}
\partial_z ( {\bf\partial\cdot A}- \partial_z A_t)-2\Im(\Phi^* \partial_z \Phi)  = 0,&
\end{eqnarray}
where $\Im(.)$ represents imaginary part, $\square=\partial_x^2  + \partial_y^2 $ and ${\bf\partial\cdot A}=\partial_x A_x + \partial_y A_y $. 
The above five equations are not independent. Indeed, their L.H.S. satisfy the following constraint equation
\begin{eqnarray}
-\frac{d}{dt}\text{Eq. \eqref{eom1}}-\frac{d}{dz}\text{Eq. \eqref{eom2}}+2\frac{d}{dx}\text{Eq. \eqref{eom3}}+2\frac{d}{dy}\text{Eq. \eqref{eom4}}\equiv4\Im(\text{Eq. \eqref{eompsi}}\times\Phi^*). 
\end{eqnarray}
 Hence, there are four independent equations for four fields,  $\Phi, A_t, A_x$, and $A_y$. This also implies that the choice of the gauge $A_z=0$ is feasible for the setup of the system.

\subsection{Boundary conditions}
Near the AdS boundary $z\to0$, the asymptotic expansions of the fields are  (we have set $m^2= -2$)
\begin{eqnarray}
A_\mu\sim a_\mu+b_\mu z+\dots,~~~\Psi=z\left(\Psi_0+\Psi_1 z+\dots\right)
\end{eqnarray}
From AdS/CFT dictionary, $a_t, a_i~ (i=x, y)$ and $\Psi_0$ are interpreted as the chemical potential, potentials related to turning on electric fields, and source of scalar operators on the boundary, respectively. Their corresponding conjugate variables can be evaluated by varying the renormalized on-shell action $S_{\rm ren}$ with respect to these source terms. 

From holographic renormalization~\cite{Skenderis02}, the counter-term of the scalar field is $S_{\rm ct}=\int d^3x\sqrt{-\gamma}\Psi^*\Psi$, where $\gamma$ is the determinant of the reduced metric on the $z\to0$ boundary.  We imposed the Neumann boundary conditions for the gauge fields as $z\to0$, in order to get the dynamical gauge fields in the boundary~\cite{Domenech10}. Therefore, the surface term of gauge fields $S_{\rm surf}=\int d^3x\sqrt{-\gamma}n^\mu F_{\mu\nu}A^\nu$ near the boundary $z\to0$ should be added in order to have a well-defined variation. Here, $n^\mu$ is the normal vector perpendicular to the boundary. Hence, by adding the counter-term $S_{\rm ct}$ and surface term $S_{\rm surf}$ into the divergent on-shell action,  one can achieve the finite renormalized on-shell action $S_{\rm ren}$. 

Varying $S_{\rm ren}$ with respect to the scalar source $\Psi_0$, one can obtain the expectation value of the order parameter as  $\langle O\rangle=\Psi_1$. Expanding the $z$-component of Maxwell equations near boundary $z\to0$ yields $\partial_tb_t+\partial_iJ^i=0$, which is a conservation equation of the charge density and current on the boundary $z\to0$. From the variation of $S_{\rm ren}$ one finds that $b_t=-\rho$ with $\rho$ the charge density and $J^i=-b_i-(\partial_ia_t-\partial_ta_i)$, which is the current along $i$-direction.

On the boundary $z\to0$, we set $\Psi_0=0$ in order to have a spontaneous symmetry-breaking of the order parameter. The Neumann boundary conditions for the gauge fields are imposed by satisfying the above conservation equations. Therefore, dynamical gauge fields on the boundary can be achieved and result in the spontaneous formation of magnetic vortices. Moreover, we impose the periodic boundary conditions for all the fields along the $(x, y)$-directions. Near the horizon $z_h$ we set a vanishing boundary condition for $A_t$ and the regular finite boundary conditions for other fields. According to the dimensional analysis, the temperature of the black hole $T$ has mass dimension one, while the mass dimension of the charge density $\rho$ is two. Therefore, $T/\sqrt{\rho}$ is a dimensionless quantity. From holographic superconductor~\cite{Hartnoll08}, decreasing the temperature is equivalent to increasing the charge density. Therefore, in order to quench the temperature linearly according to KZM, we actually quench the charge density $\rho$ as $\rho(t)=\rho_c\left(1-t/\tau_Q\right)^{-2}$ with $\rho_c\approx4.06$, which is the critical charge density of the holographic superconductor in the equilibrium state.

\subsection{Numerical schemes}  
As discussed in the main text, the system is thermalized before quenching in order to have an equilibrium initial state in the normal phase. During the quenching, we evolve the system by using the fourth-order Runge-Kutta (RK4) method with a time step $\Delta t=0.02$. In the radial $z$-direction, we adopt the Chebyshev pseudo-spectral method with 21 grid points~\cite{Trefethen00book}. Since in the $(x, y)$-directions, all the fields are periodic, we take advantage of the Fourier decomposition along the  $(x, y)$-directions with the spatial spacing $\Delta x=\Delta y=0.25$. The codes are indeed very robust to the grid points in the time and spatial directions. We choose these grid points in our paper considering both the computational time and the quality of the numerical results. Filtering of the high momentum modes is implemented following the ``$2/3$'s rule'', with the uppermost one-third Fourier modes being removed~\cite{CheslerYaffe14}. 

\end{document}